\documentclass[preprint2]{aastex63}

\accepted{September 18, 2020}
\submitjournal{ApJL}

\shorttitle{Rays from IC 5063}
\shortauthors{Maksym et al.}

\usepackage[applemac]{inputenc}
\usepackage{hyperref,breakurl}
\def\UrlBreaks{\do\/\do-}
\usepackage{ifthen,graphicx}
\usepackage{epstopdf,placeins}
\usepackage{soul,color,amsmath}
\usepackage{epstopdf,overpic}
\usepackage{soul,color,amsmath}
\def\mathnew{\mathsurround=0pt}
\def\simov#1#2{\lower 2.5pt\vbox{\baselineskip0pt \lineskip-.5pt
\ialign{$\mathnew#1\hfil##\hfil$\crcr#2\crcr\sim\crcr}}}
\def\simless{\mathrel{\mathpalette\simov <}}
\def\simgreat{\mathrel{\mathpalette\simov >}}
\newcommand{\MeV}{Me\kern-0.11em V}
\newcommand{\keV}{ke\kern-0.11em V}

\newcommand{\z}{{\it z\/}}

\newcommand{\kms}{\ensuremath{\rm km\;s}^{-1}}

\newcommand{\yellow}[1]{\textcolor{yellow}{#1}}
\newcommand{\white}[1]{\white{white}{#1}}
\makeatletter
\newcommand\iont[2]{{#1$\;${\small\expandafter\@slowromancap\romannumeral #2@\relax}}}
\makeatother
\makeatletter
\newcommand{\raisemath}[1]{\mathpalette{\raisem@th{#1}}}
\newcommand{\raisem@th}[3]{\raisebox{#1}{$#2#3$}}
\makeatother
 \makeatletter \g@addto@macro\UrlSpecials{\do\!{\newline}}
 
\begin{document}

\title{Crepuscular Rays from the Highly Inclined Active Galactic Nucleus in IC 5063\footnote{Based on observations made with the NASA/ESA Hubble Space Telescope, obtained from the data archive at the Space Telescope Science Institute. STScI is operated by the Association of Universities for Research in Astronomy, Inc. under NASA contract NAS 5-26555. These observations are associated with programs \#15444 and \#15609.}}

\correspondingauthor{W. Peter Maksym; @StellarBones}
\email{walter.maksym@cfa.harvard.edu}

\author[0000-0002-2203-7889]{W. Peter Maksym}
\affiliation{Center for Astrophysics \textbar\ Harvard \& Smithsonian, 60 Garden St., Cambridge, MA 02138, USA}

\author[0000-0002-2617-5517]{Judy Schmidt}
\affiliation{Astrophysics Source Code Library, University of Maryland, 4254 Stadium Drive, College Park, MD 20742, USA}

\author[0000-0002-6131-9539]{William C. Keel}
\affiliation{Department of Physics and Astronomy, University of Alabama, Box 870324, Tuscaloosa, AL 35487, USA}

\author[0000-0002-3554-3318]{Giuseppina Fabbiano}
\affiliation{Center for Astrophysics \textbar\ Harvard \& Smithsonian, 60 Garden St., Cambridge, MA 02138, USA}

\author[0000-0002-3365-8875]{Travis C. Fischer}
\affiliation{Space Telescope Science Institute,
3700 San Martin Drive, Baltimore, MD 21218, USA}

\author[0000-0001-7516-4016]{Joss Bland-Hawthorn}
\affiliation{Sydney Institute of Astronomy, School of Physics, University of Sydney, Australia}
\affiliation{ARC Centre of Excellence for All Sky Astrophysics in 3D, Australia}

\author[0000-0002-3026-0562]{Aaron J. Barth}
\affiliation{Department of Physics and Astronomy, University of California at Irvine, 4129 Frederick Reines Hall, Irvine, CA 92697-4575, USA}

\author[0000-0001-5060-1398]{Martin Elvis}
\affiliation{Center for Astrophysics \textbar\ Harvard \& Smithsonian, 60 Garden St., Cambridge, MA 02138, USA}

\author[0000-0002-0616-6971]{Tom Oosterloo}
\affiliation{ASTRON, The Netherlands Institute for Radio Astronomy, Postbus 2, 7990 AA Dwingeloo, The Netherlands}
\affiliation{Kapteyn Astronomical Institute, University of Groningen, Postbus 800, 9700 AV Groningen, The Netherlands }

\author[0000-0001-6947-5846]{Luis C. Ho}
\affiliation{Kavli Institute for Astronomy and Astrophysics, Peking University, Beijing 100871, People’s Republic of China}
\affiliation{Department of Astronomy, School of Physics, Peking University, Beijing 100871, People’s Republic of China}

\author[0000-0002-3560-0781]{Minjin Kim}
\affiliation{Department of Astronomy and Atmospheric Sciences, College of Natural Sciences, Kyungpook National University, Daegu 41566, Republic of Korea}

\author{Hyunmo Hwang}
\affiliation{Department of Astronomy and Atmospheric Sciences, College of Natural Sciences, Kyungpook National University, Daegu 41566, Republic of Korea}

\author{Evan Mayer}
\affiliation{Independent researcher, Tucson, AZ 85719, USA}



\begin{abstract}

On Earth near sunset, the sun may cast ``crepuscular rays" such that clouds near the horizon obscure the origin of light scattered in bright rays.  In principle, AGN should be able to produce similar effects.  Using new {\it Hubble Space Telescope} ({\it HST}) near-infrared and optical observations, we show that the active galaxy IC 5063 contains broad radial rays extending to  $\simgreat11$\,kpc from the nucleus.  We argue that the bright rays may arise from dusty scattering of continuum emission from the active nucleus, while the dark rays are due to shadowing near the nucleus, possibly by a warped torus.  We also consider alternative AGN-related and stellar origins for the extended light.

\end{abstract}

\keywords{AGN host galaxies (2017), Interstellar dust (836), Interstellar scattering (854), Galaxy mergers (608), Galaxy winds (626), Galaxy morphlogy (582)}

\section{Introduction} \label{sec:intro}

Crepuscular rays are a commonly observed atmospheric phenomenon in which clouds near the horizon obscure the origin of sunlight which is scattered in the Earth's atmosphere.  The scattered sunlight appears as bright rays which point back towards the hidden light source, and the contrast between the dark (shadowed) rays and bright (scattered) rays is most obvious during twilight \citep{Minnaert74}.  This phenomenon can also be seen in an astronomical objects such as NGC 2261 (``Hubble's Nebula") where clouds of dust obscure emission from the star R Mon, and active galactic nuclei (AGN) should also be able to produce similar effects in principle.

Unlike many AGN, IC 5063 displays powerful kpc-scale radio outflows which are oriented directly into the plane of the galaxy, colliding with the nuclear interstellar medium (ISM; \citealt{Morganti98,Schmitt03,Morganti15,Dasyra15,Oosterloo17}).  Models of these jet-ISM interactions by \cite{Mukherjee18} broadly replicate the key CO features found by \cite{Morganti15} with ALMA.  These models also predict venting of hot plasma perpendicular to the disk and entrainment of cooler clumps and filaments in the halo, which may be carried to large radii \citep{McCourt15,BB16,Gronnow18}.

The proximity of IC 5063 ($z=0.01140$; $D\sim47.9$\,Mpc) and its highly obscured (log\,$[n_{\rm H}/{\rm cm}^{-2}]=23.55$;  \citealt{Ricci17}) nucleus allow us to spatially resolve bright extended narrow-line emission that is associated with the radio outflows in the optical \citep{Morganti98,Schmitt03} and X-rays \citep{GG17}.  

Under the standard AGN model \citep{UP95}, the torus comprises a dusty, ring-like structure that extends perpendicularly to the bicone and obscures emission that arises close to the black hole and accretion disk.  At larger scales, nuclear starbursts and the galactic disk may also contribute alignment-dependent obscuration \citep{HA18}.  In IC 5063, the misalignment of the jet with the galactic disk suggests that obscuration by the torus could occur preferentially out of the galactic plane.  

By examining new broadband, continuum-dominated {\it HST} observations spanning red and near-infrared wavelengths, J. Schmidt detected\footnote{\url{https://twitter.com/SpaceGeck/status/1201350966945017856}}  a series of bright and dark rays extending from the nucleus of IC 5063.   
In this paper, we establish the presence of these features, examine their properties in detail, and investigate their possible origins.
 Current AGN models and observations \citep{Elvis00, HA18} typically entail wavelength-dependent stratification of radiative transmission as a function of the angle from the nuclear axis.  These effects are typically easier to see at sub-kpc scales \citep[e.g.][]{Maksym16,Maksym17,Maksym19} due to the presence of dense ISM where the plane of the galaxy  intersects with the bicone.  For such effects to be seen at larger distances, some kind of ``screen" is necessary to reprocess the nuclear emission \citep{Keel15,Keel17}.  The unusual configuration of IC 5063 may provide such a combination in the near-infrared.

As in \cite{Oosterloo17}, we adopt an angular size distance of 47.9 Mpc and a scale of 1\arcsec = 232 pc for IC 5063.

\begin{figure*}
\centering

\hspace{0.09in}\raisebox{0.255cm}[0pt][0pt]{\begin{overpic}[width=0.463\textwidth,trim={0cm 0cm 0cm 0.cm},clip]{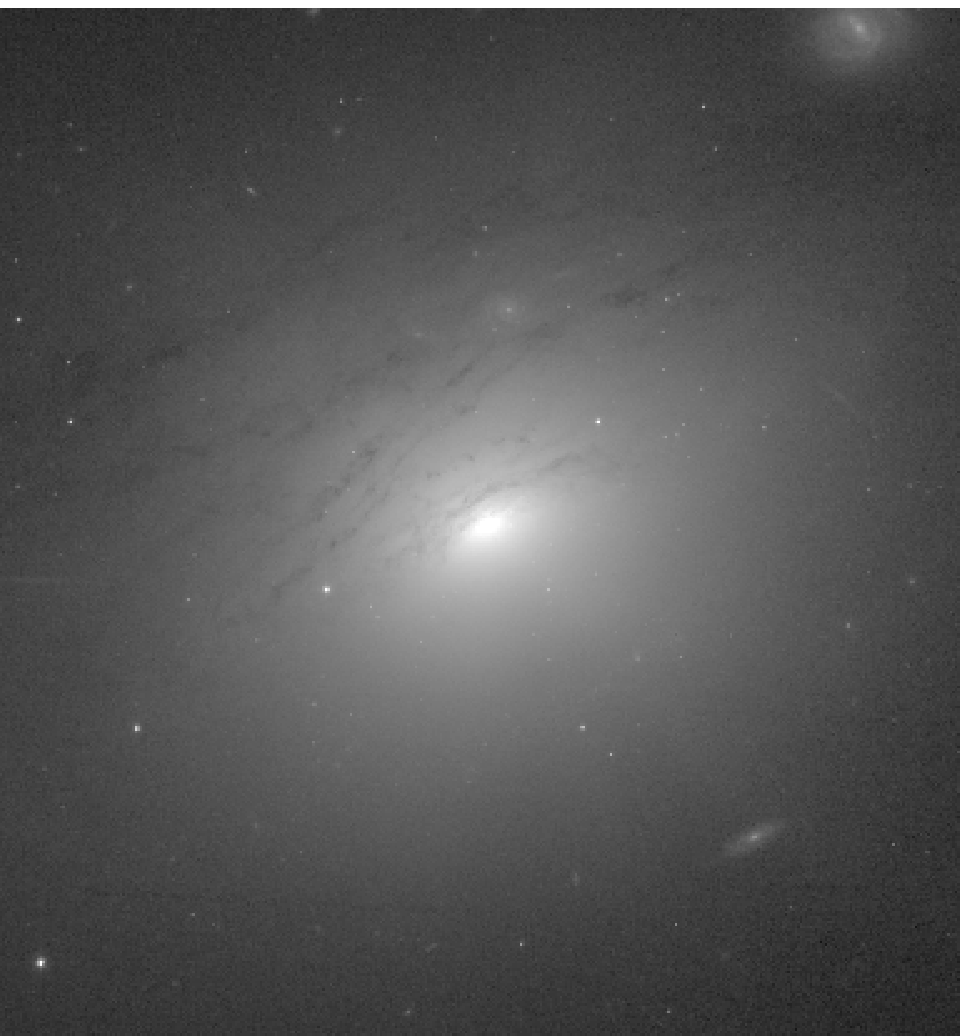}
	\put(4,88){\color{white}\rule{2.3cm}{0.8cm}}
	\put(5,91.5){{\parbox{5.5cm}{%
			\LARGE F814W
			}}}
\end{overpic}}\hspace{0.1in}
\begin{overpic}[width=0.489\textwidth,trim={0cm 0cm 0 0cm},clip]{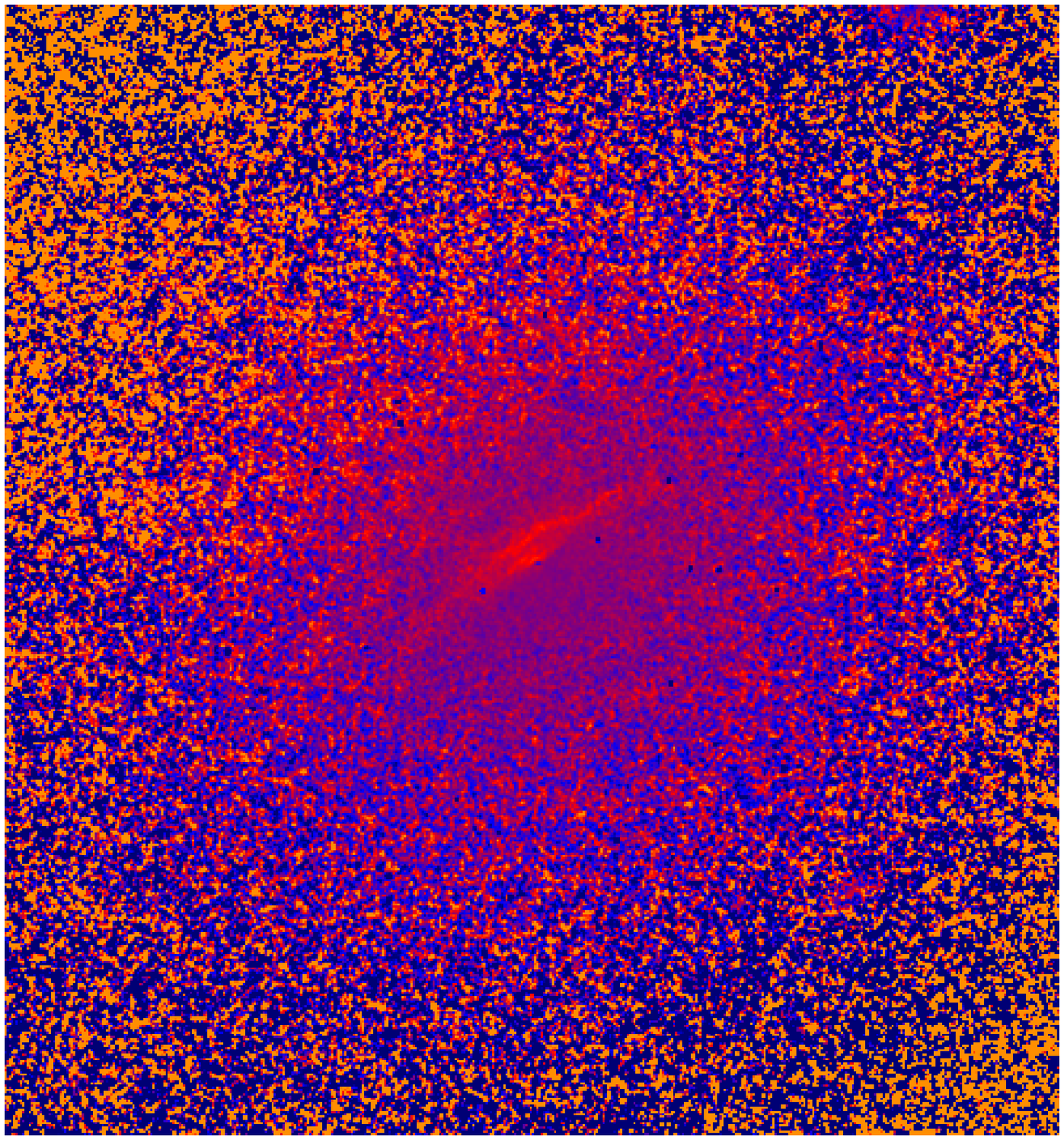}
	\put(4,86){\color{white}\rule{4.8cm}{0.8cm}}
	\put(5,89){{\parbox{5.5cm}{%
			\LARGE F763M/F814W
			}}}
\end{overpic}\hfill\null
\begin{overpic}[width=0.4895\textwidth]{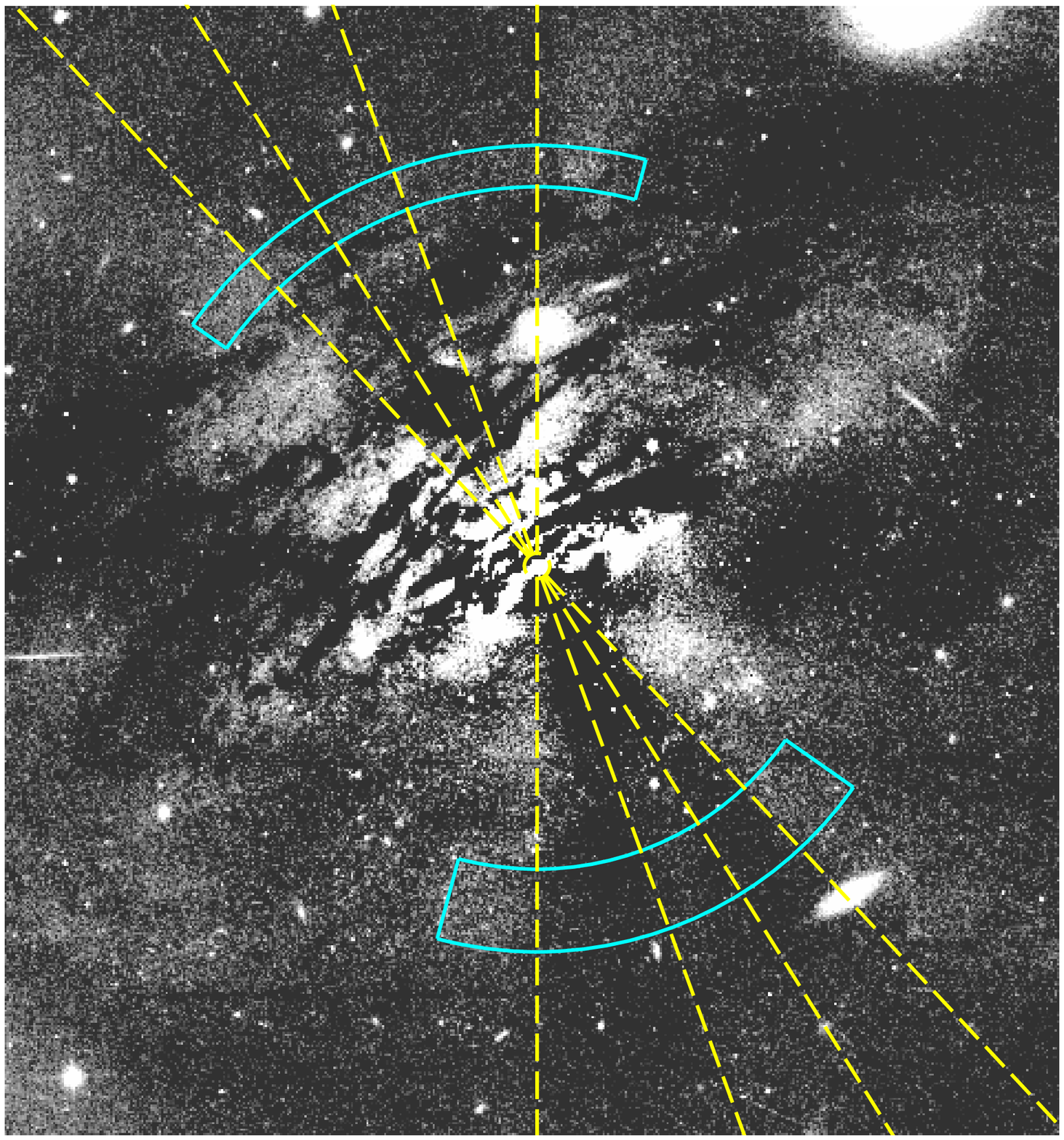}
	\put(5,88){\color{white}\rule{5.7cm}{0.7cm}}
	\put(5,91){{\parbox{6.0cm}{%
			\large F814W: model-subtracted
			}}}
	\put(19,81){{\rotatebox{315}{%
			\yellow{\bf dark ray}
			}}}
	\put(38,87){{\rotatebox{280}{%
			\yellow{\bf dark ray}
			}}}
	\put(48,27){{\rotatebox{280}{%
			\yellow{\bf dark ray}
			}}}
	\put(64,25){{\rotatebox{305}{%
			\yellow{\bf dark ray}
			}}}
\end{overpic}
\begin{overpic}[width=0.4895\textwidth]{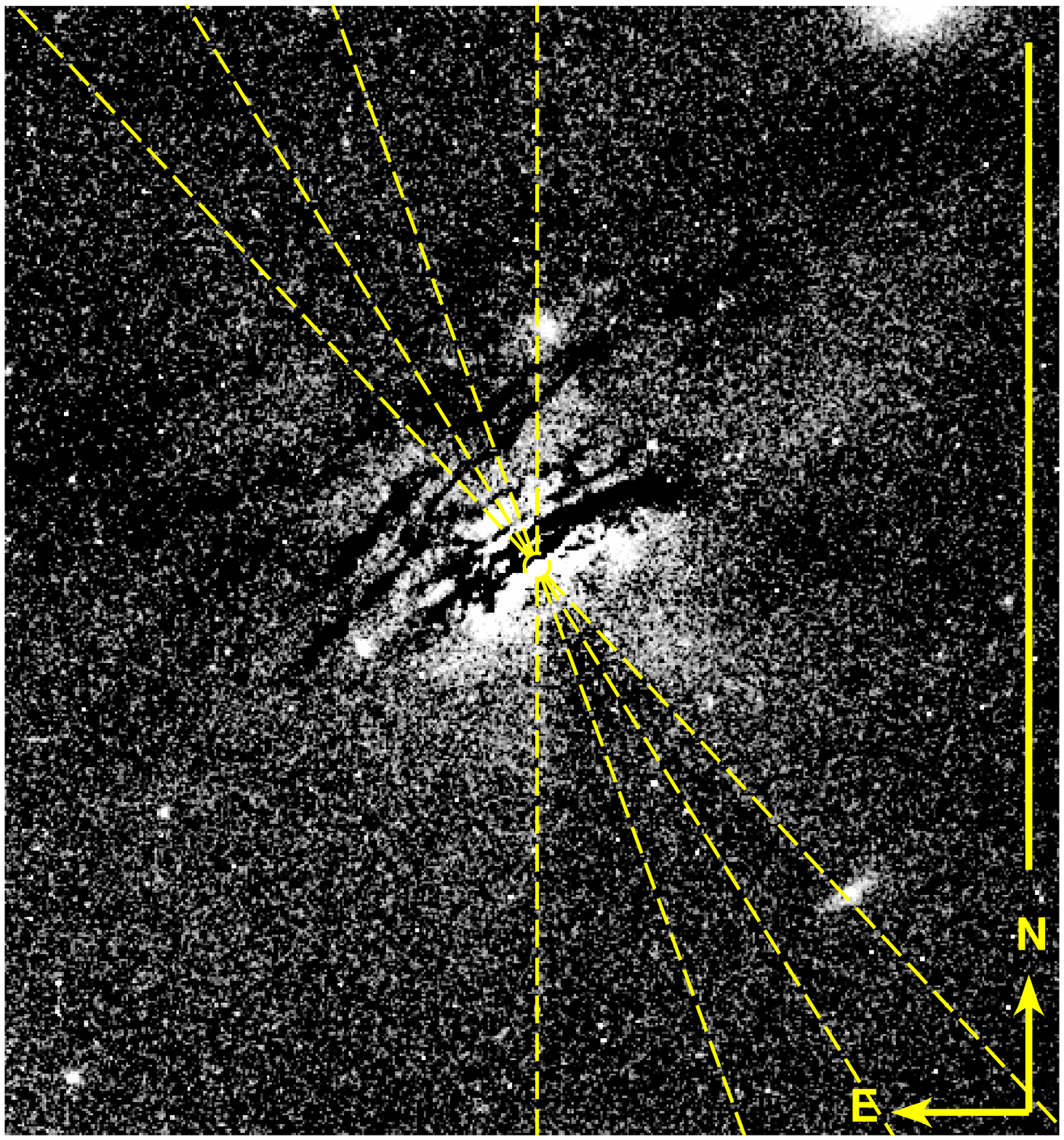}
	\put(5,88){\color{white}\rule{5.7cm}{0.7cm}}
	\put(5,91){{\parbox{6.0cm}{%
			\large F763M: model-subtracted
			}}}
	\put(19,81){{\rotatebox{315}{%
			\yellow{\bf dark ray}
			}}}
	\put(38,87){{\rotatebox{280}{%
			\yellow{\bf dark ray}
			}}}
	\put(48,27){{\rotatebox{280}{%
			\yellow{\bf dark ray}
			}}}
	\put(64,25){{\rotatebox{305}{%
			\yellow{\bf dark ray}
			}}}
	\put(83,48){{\rotatebox{90}{%
			\large\yellow{\bf 1 arcmin}
			}}}
\end{overpic}
\caption{Continuum imaging of IC 5063.  Scale and orientation indicated in bottom right image.  {\bf Top Left:} Processed, unenhanced F814W image.  {\bf Top Right:} Color map from the image ratio F763M/F814W, smoothed with a 0\farcs 25 boxcar.  This ratio map lacks well-defined large-scale features, except for the nuclear dust lane.  The color scale is chosen to enhance similar features.
{\bf Bottom Left:} Ellipsoidal models are fit to F814W and subtracted.  Both dust lanes and dark/light `rays' are now prominent.  Dashed yellow lines mark the edges of dark rays to SW and NE of nucleus, and the bright ray at the center of each dark zone.  Regions extracted for azimuthal profiles (see Fig. \ref{fig:az}) are marked with cyan arcs. {\bf Bottom Right:} Dark/light `rays' remain prominent in the F763M continuum filter, despite narrower bandwidth and shorter exposure.}
\label{fig:rays-grid}
\end{figure*}

\begin{figure*}
\centering
\begin{overpic}[width=0.99\textwidth]{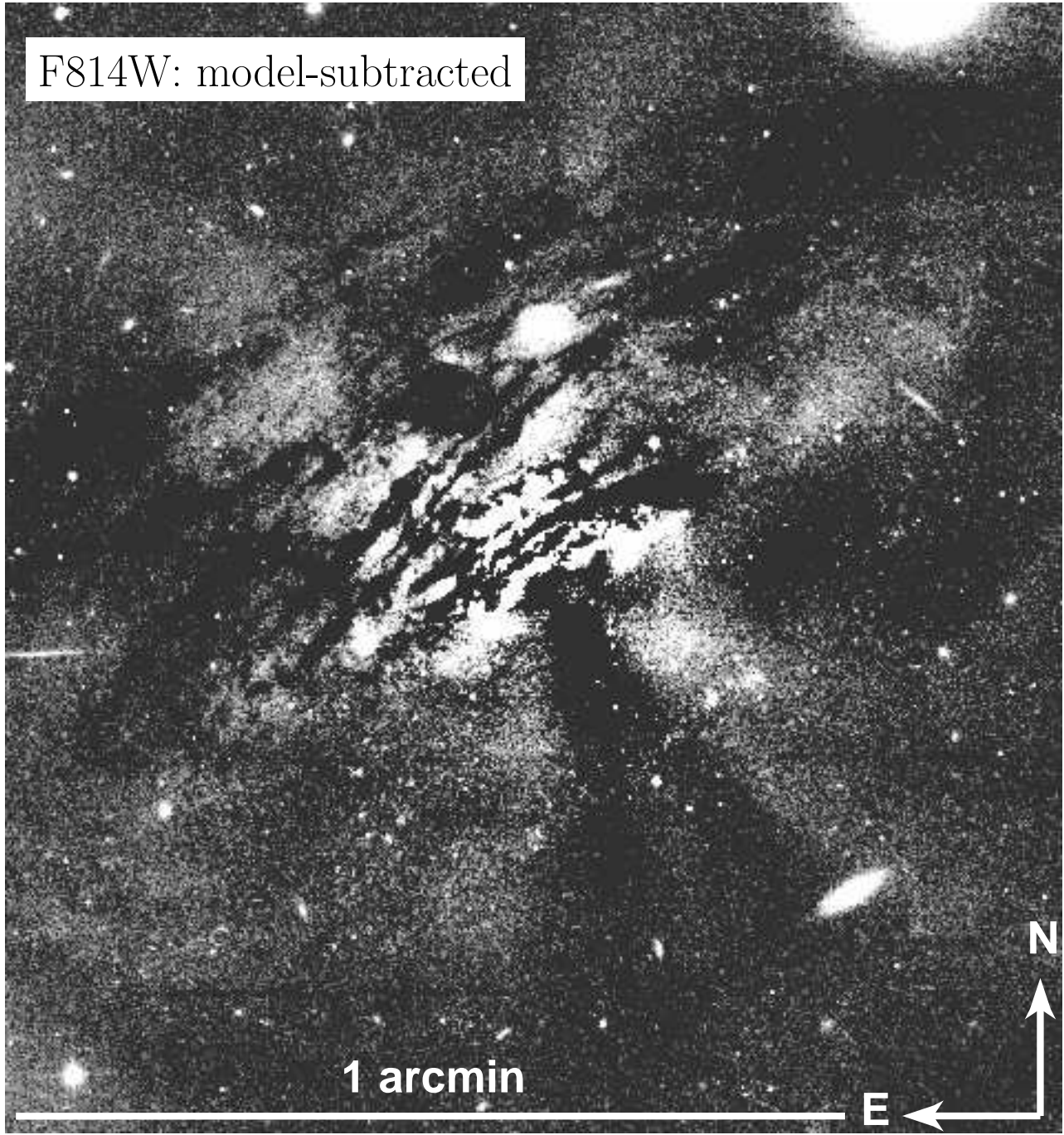}
	\put(4,89){\color{white}\rule{8.cm}{1.0cm}}
	\put(5,91){{\parbox{9.5cm}{%
			\LARGE F814W: model-subtracted
			}}}
\end{overpic}
\caption{As with Fig. \ref{fig:rays-grid}, bottom left.  The image has been magnified and labels have been removed to enable viewing without interference.}
\label{fig:rays-big}
\end{figure*}

\section{Observations and Data}

IC 5063 was observed by {\it HST} on 2019 November 25 as part of a snapshot survey of nearby AGN host galaxies (Program ID 15444, PI: Barth). This survey was one of three programs selected through the special 2017 mid-cycle ``gap-filler'' snapshot proposal call. The sample for Program 15444 consists of AGN at $z<0.1$ selected from the {\it Swift}-BAT AGN sample \citep{Koss17} that had no prior {\it HST} imaging in broad {\it I}-band filters. IC 5063 was observed with the ACS WFC and F814W filter with two dithered exposures for a total exposure time of 674\,s.  At the redshift of IC 5063, F814W primarily covers the NIR stellar continuum centered on $\lambda_\mathrm{pivot}=8045\,$\AA ($>1$\% transmission between $\lambda\lambda6868,9626\,$\AA).   IC 5063 was also observed for 307\,s on 2019 March 7 (Program ID 15609, PI Fabbiano) in the F763M filter ($\lambda_\mathrm{pivot}=7615\,$\AA, $>1$\% between $\lambda\lambda7165,8092\,$\AA).  All filter wavelengths are in the observer frame. 

Data were reduced using the {\it HST} data processing package AstroDrizzle \citep{astrodrizzle} according to standard techniques.  Cosmic rays were removed via L.A. Cosmic \citep{lacosmic}.   All instrumental conversions to flux are inferred from standard {\it HST} filter values.

For direct comparison between the F763M and F814W bands, we sampled them to an identical pixel scale and produced a color map by calculating the ratio between the two images.

To enhance the radial features found by J. Schmidt, we performed ellipse-fit modeling to both the F814W and F763M images according to the algorithm used by \cite{Jedrzejewski87},  as implemented in the IRAF {\tt ellipse} task.  The best-fit 2-D models were then subtracted from the images.  This process removes most of the smooth galaxy light and improves the contrast of the radial features.   Such an approach faithfully retains sharp radial edges but not smooth transitions.

The processed image, color ratio map and ellipsoid-subtracted versions are displayed in Fig. \ref{fig:rays-grid} and \ref{fig:rays-big}.

Before producing radial profiles of the diffuse extended emission, we used the {\it ciao} tools {\tt wavdetect} and {\tt dmfilth} to identify and excise bright interloper objects in the ellipsoid-subtracted F814W image.  Additional objects were identified by eye and removed.

\vspace{10mm}
\section{Results}

The dark rays identified by J. Schmidt in the ACS F814W image become prominent when elliptical models of the host starlight are subtracted, as seen in Figs. \ref{fig:rays-grid} and \ref{fig:rays-big}.  They extend $\simgreat50\arcsec$ from the nucleus to both NE and SW, and each subtends an angle of $\sim43\degr$ (counter-clockwise [CCW] from N and S respectively).  The rays are also present in the WFC3 F763M image.  We do not identify any obvious features in the F763M/F814W ratio map except for the major dust lanes.  The two prominent edges on each side intersect within $\sim2$\arcsec\ of the brightest nuclear feature.

The ellipse-fitting includes the dark sectors as well as dust lanes and bright features in its masked regions, and may therefore fail to enhance smooth transitions.  It will faithfully reproduce the location of the relatively sharp edges defining the dark sectors but gives essentially no information on whether exterior bright zones have fuzzy edges towards the galaxy major axis or whether they form complete wedges.  Typical contrast across the dark edges is $\sim5\%$ of the galactic light intensity. 

\begin{figure} 
\centering
\noindent
\includegraphics[width=0.99\linewidth]{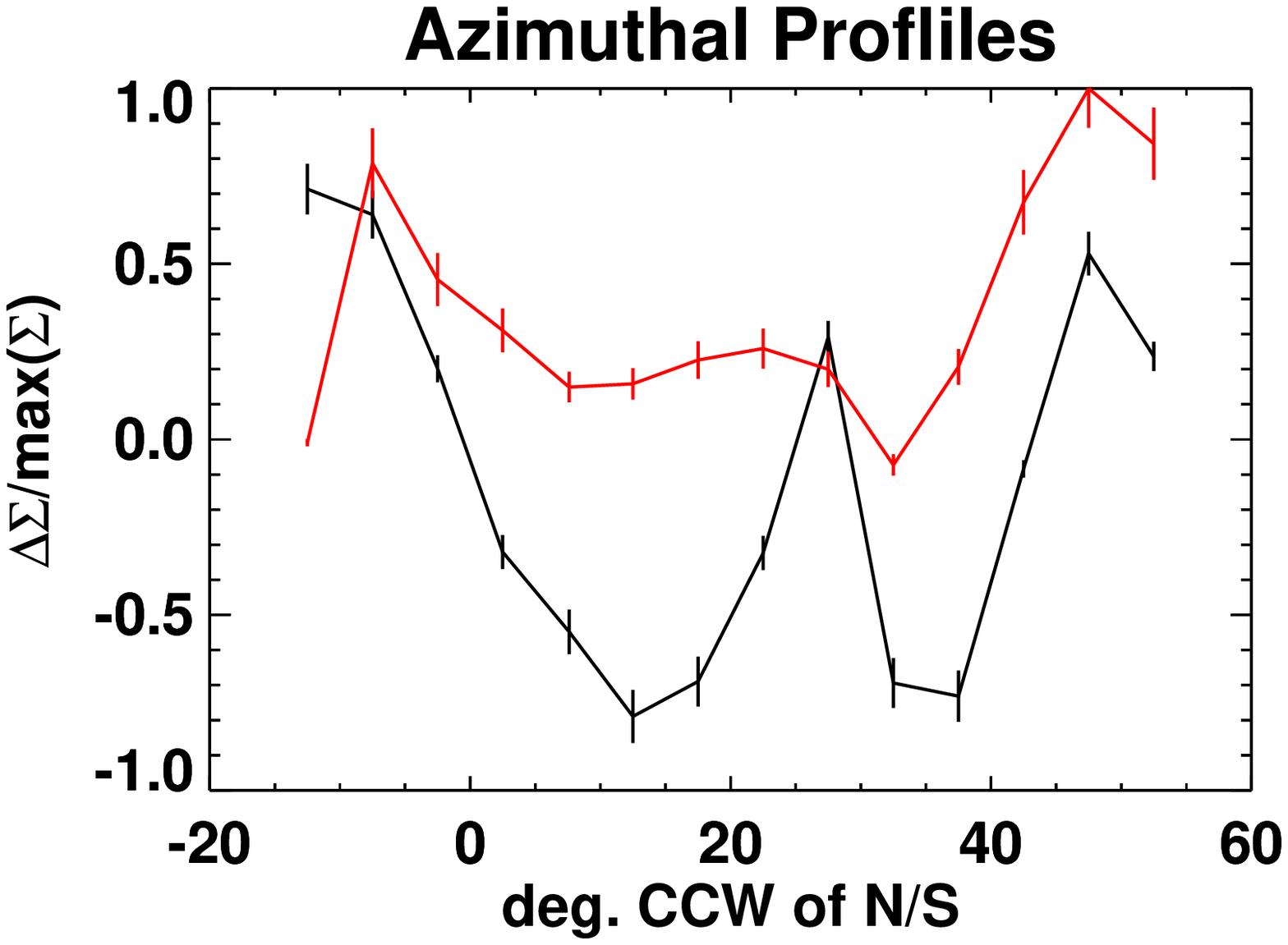}\par
\includegraphics[width=0.99\linewidth]{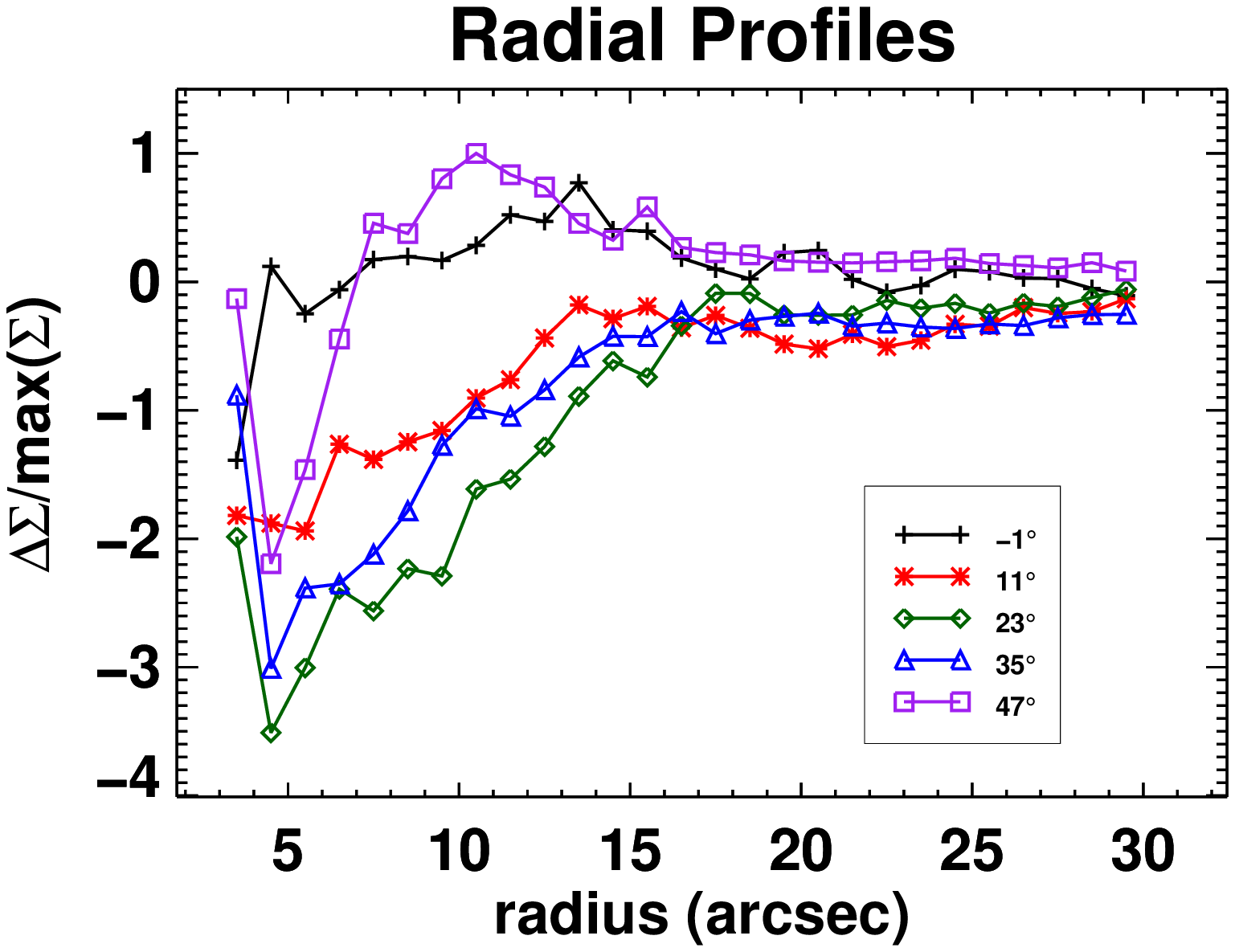}\par
\caption{{\bf Top:} Azimuthal surface brightness profiles taken from cross sectional arcs, as depicted in the ellipsoid-subtracted F814W image in Fig. \ref{fig:rays-grid} (lower-left).  Black points are extracted from annular segments at [$22\farcs 0<r<28\farcs 0$] from the nucleus, and  the X-axis denotes degrees CCW from south.  For red points, the X-axis denotes degrees CCW from north, and [$22\farcs 0<r<28\farcs 0$].  The Y-axis is the local deviation from the model, renormalized by the maximum extracted value in this plot.  {\bf Bottom:} Radial surface brightness plots also using the ellipsoid-subtracted F814W image in Fig. \ref{fig:rays-grid} (lower-left).  Angles designate the midpoint orientation of the wedge, measured CCW from south.  Profiles are taken from annular arcs of uniform 12$\degr$ span and 1\arcsec\ thickness.  As above, Y-axis is the local surface brightness deviation from the model, renormalized by the maximum extracted value in this plot.  Error bars are comparable to data point sizes.
} 
\label{fig:az}
\end{figure}

In order to measure the relative strength of the light and dark rays, we extracted azimuthal profiles with $5\degr$ azimuthal bins (Fig. \ref{fig:az}) from the ellipsoid-subtracted F814W images (Fig. \ref{fig:rays-grid}) across the dark sectors at [$-20\degr<\phi<60\degr$] relative to both north and south.  We selected radial ranges devoid of both bright sources and strong planar dust lanes.  Dust lanes cover more of the NE dark sector, so the NE annulus has only $50\%$ of the thickness of the SW annulus.  We see in Fig. \ref{fig:az} (top panel) that the dark sectors extend from $\sim-5\degr$ to $\sim45\degr$ CCW of [N, S] for the [NE, SW] zones.  We also confirm the presence of a central bright ray at $\sim25\degr-30\degr$ which has $\sim50\%$ surface brightness relative to the bright edges that bound the dark rays.  This bright ray is both narrower and better-defined in the SW cone.

We also extracted radial profiles for five 12$\degr$ wedges spanning $3\arcsec<r<30\arcsec$ from the nucleus (Fig. \ref{fig:az}, bottom panel).  These correspond roughly to the two SW dark dark rays ($\theta=[11\degr, 35\degr]$ measured CCW from south), the bright rays that bound them ($\theta=[-1\degr, 47\degr]$), and the central bright ray ($\theta=23\degr$).  The contrast between bright and dark rays is evident at $5\arcsec<r<28\arcsec$, but the central bright ray does not become evident until $r\simgreat16\arcsec$, which corresponds roughly to tangential features that resemble clumpy stellar streams.  At larger radii, all five profiles are comparatively flat, although the bright outer rays ($\theta=[-1\degr, 47\degr]$) do decay with radius.

A  bright background galaxy at ($\alpha,\delta$)=(20$^{\rm h}$\,51$^{\rm m}$\,59$^{\rm s}$, -57$\degr$04\arcmin 31\farcs 7) straddles the edge of the SW dark sector.  In order to measure color differences across the edge of the dark cones which may result from extinction, we compare F763M/F814W flux ratios on the NW and SE sides of this galaxy.  We find the respective ratios to be comparable at the level of uncertainty, $\sim1\%$.

\section{Discussion}

\begin{figure*} 
\noindent
\hspace{0.05\textwidth}
\begin{overpic}[width=0.33\textwidth]{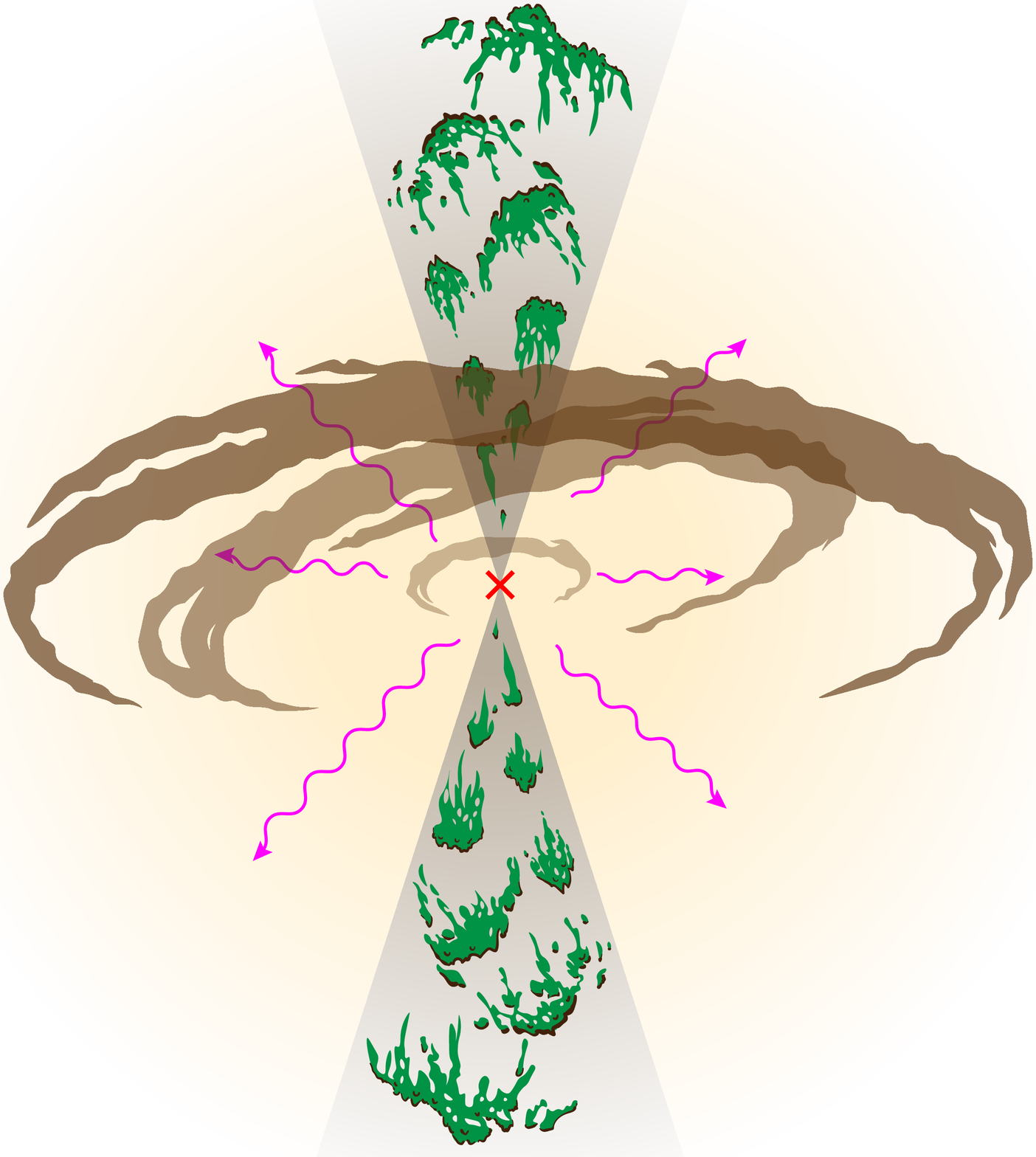}
	\put(65,4){{\parbox{6cm}{%
			\bf\small Dusty Cone
			}}}
\end{overpic}
\hspace{0.1\textwidth}
\begin{overpic}[width=0.43\textwidth]{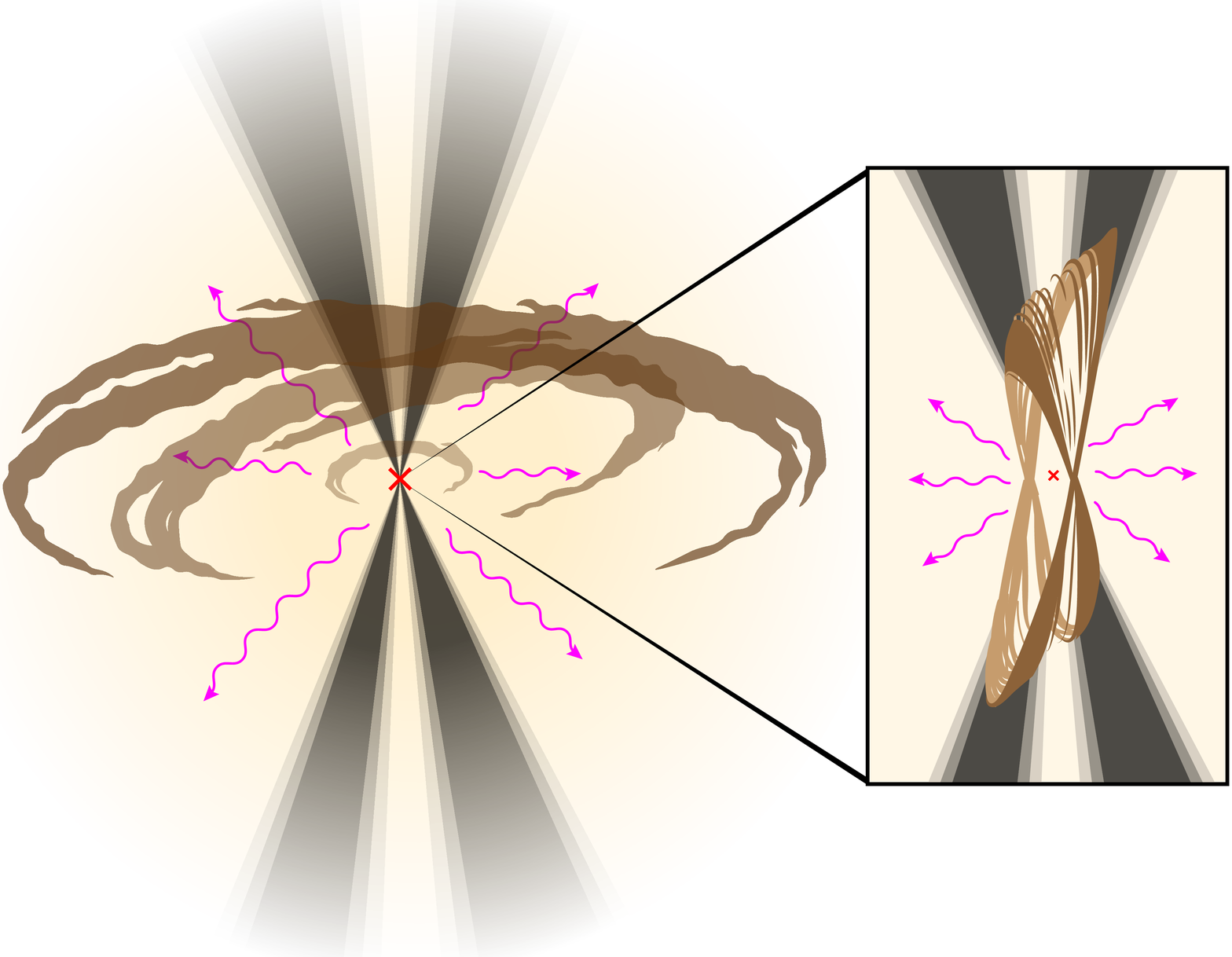}
	\put(65,5){{\parbox{6cm}{%
			\bf\small Shadow of a\\  Warped Torus
			}}}
\end{overpic}\par
\vspace{0.2in}
\hspace{0.05\textwidth}
\begin{overpic}[width=0.33\textwidth]{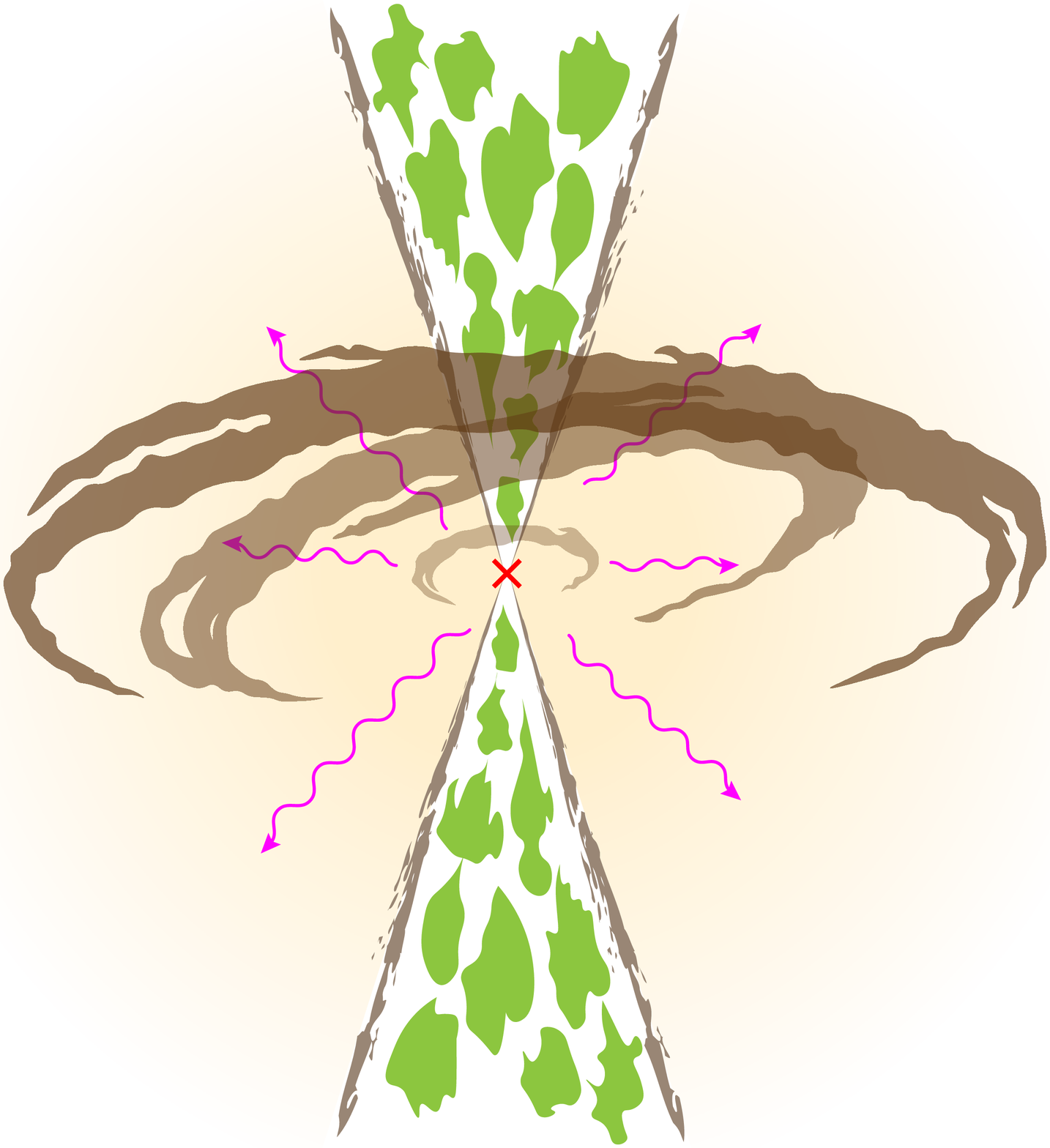}
	\put(70,3){{\parbox{6cm}{%
			\bf\small Vacated Cone
			}}}
\end{overpic}
\hspace{0.12\textwidth}
\raisebox{0.6cm}[0pt][0pt]{
\begin{overpic}[width=0.28\textwidth]{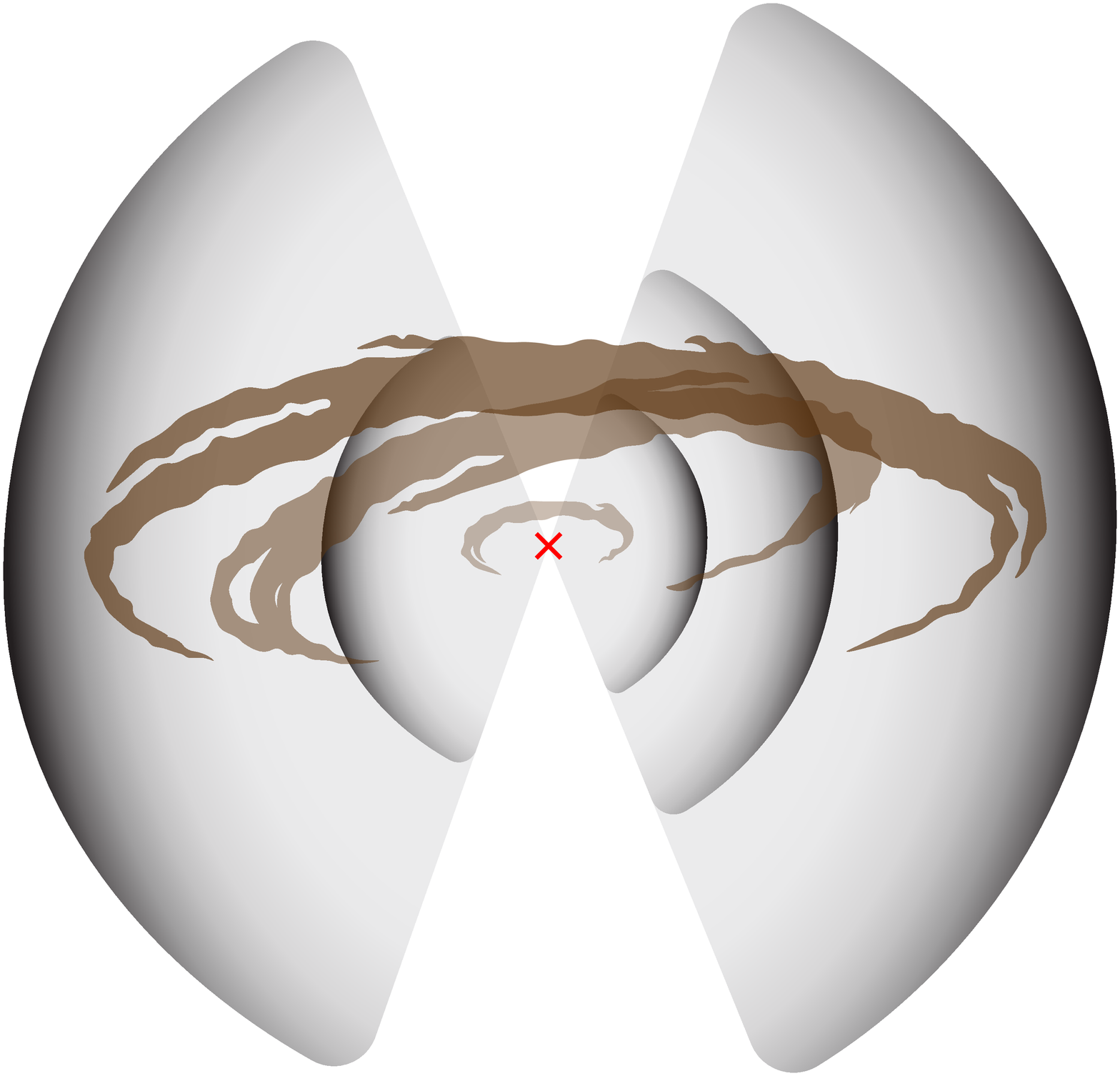}
	\put(90,-8){{\parbox{6cm}{%
			\bf\small Stellar Fans
			}}}
\end{overpic}
}
\caption{{\bf Cartoons to represent possible physical scenarios:} Dust is brown or tan.  The red 'x' marks the nucleus.  {\bf 1) Upper Left:}   The diffuse excess light is caused by dusty reflection of AGN light at broad angles.  The `dark rays' contain dust (brown) entrained and circulated by hot outflows (green).  This dust may absorb radiation both radially from the AGN and in the observer's line-of-sight. 
 {\bf 2) Upper Right:}  Again, diffuse light is caused by dusty reflection of AGN emission.  The `dark rays' are caused by shadowing near the nucleus.  One possible explanation for the central `bright ray' is that shadowing is caused by warped, clumpy disk which permits some transmission at limited angles.  {\bf 3) Lower Left:}   Diffuse light is caused by dusty reflection of AGN emission. Hot outflows (light green) have vacated or destroyed the dust in the `dark rays' (white), which therefore cannot scatter AGN emission to the observer.  {\bf 4) Lower Right:} Dark rays mark the edges of post-merger stellar light.  Tidal debris fans out to form a shell galaxy, but shell formation (black) is interrupted, obscured or evolved, and is therefore not observed.  Extraplanar dust reflects little light.
} 
\label{fig:cartoons}
\end{figure*}

The most spectacular ionization cones in Seyfert 2 galaxies are commonly inclined towards the plane \citep[e.g., NGC 1068 and Circinus; see][]{VBH97}.  This configuration is well-established in IC 5063.  The radio continuum runs along the dust lane, and is associated with optically thin outflows of atomic and molecular gas at $\sim800\,\kms$ which may drive a lateral outflow \citep{Dasyra15,Oosterloo17}.  

The dark rays extending NW-SE from the nucleus in continuum bands (F763M, F814W) extend to $\simgreat11\,$kpc, which requires a galactic-scale phenomenon connected to the nucleus.  \cite{Colina91} suggest that the galaxy-scale X-shaped ionization structures seen in their ground-based [\ion{O}{3}] and H$\alpha$ maps could be explained by shadowing of the ionization cone by nuclear dust lanes and postulate the presence of a minor axis dust ring as a possible origin for the edge of the bicone.  Hints of the galaxy-scale rays seen in the {\it HST} data are visible in dust maps made by \cite{Colina91} from ground-based {\it V} and {\it R} imaging, but these were not specifically addressed.  {\it HST} shows that the rays extend from $\simless500\,$pc (below which, planar dust may overwhelm the signature) to $\simgreat 11$\,kpc.  Such extent is not obvious in the \cite{Colina91} dust maps, suggesting {\it V} declines with radius within the rays (as might occur via scattering by diffuse dust).

Our ellipsoidal decomposition's inability to enhance smooth transitions means that it is not sensitive to whether or not any scattered contribution has uniform surface brightness all the way around to the major axis of the system.

\subsection{Scattering from Extended Dust?}

If the bright rays shown in Figs. \ref{fig:rays-grid} and \ref{fig:rays-big} are caused by scattering of light from the AGN by extended galactic dust, then the observed intensity in excess of starlight should be compatible (to first order) with the energy budget from the AGN, and with simple assumptions about optically thin dust scattering. 

Suppose $I_S\propto \eta \tau L_\mathrm{bol}r^{-2}$, where $I_S$ is the intensity of scattered flux at radius $r$, $\tau$ is the optical depth to scattering for a given path, $L_\mathrm{bol}$ is the bolometric luminosity \citep{Nicastro03}, and $\eta$ is a scaling correction that is a function of waveband, dust scattering efficiency, distribution, anisotropy, and other properties.  Let $\tau\sim\sigma_d n_d l$ where $\sigma_d=\pi(a/2)^2$ is the scattering cross-section, $a$ is a characteristic grain diameter, $n_d$ is the characteristic dust number density, and $l$ is the relevant path length.  If we take the median surface brightness of a bright ray at $r\sim27\arcsec$ ($\sim6$\,kpc) and assume that the measured monochromatic flux ($\lambda F_\lambda$) at a given projected radius is scattered from a cube subtending the WFC3 pixel solid angle, then we see that reasonable assumptions for relevant properties are compatible with the measured excess brightness for a scaling factor of $\eta\sim1$ (i.e. no modification):

\begin{equation}
\frac{1}{\eta}\sim\frac{n_d/\rm{cm}^{-3}}{8\times10^{-12}}\left(\frac{a/\rm{nm}}{100}\right)^2\frac{L_\mathrm{bol}/\rm{erg\,s}^{-1}}{8\times10^{44}}\
\end{equation}

\noindent Since $\eta$ encompasses several complex assumptions, a range of values may be reasonable.  
Also note that if the ellipsoidal models were normalized to the dark rays, the derived scattered flux could increase by a factor of a few.

There are likely multiple mechanisms for spreading dust to these high latitudes, including the merger described by \cite{Colina91}.  And lateral outflows from the disk could carry dust to large radii \citep{McCourt15,BB16,Mukherjee18,Gronnow18}, which could then obscure both galactic starlight and incident AGN radiation, possibly in a network of filamentary clouds \citep{Cooper08,Cooper09,Tanner16}.  Such a `dusty cone' scenario is depicted in Fig. \ref{fig:cartoons}, upper-left.   In principle, this dust could also obscure background galaxies.  We observed a lack of color differential across a galaxy at the edge of the SW dark sector at the $\sim1\%$ level, but this is not strongly constraining due to the intrinsic redness and band overlap of these filters.  

Such ionized lateral outflows could suppress star formation, but these continuum bands primarily trace older stars and there is no evidence of ongoing star formation in the adjacent bright rays.

If the ray patterns in the F763M and F814W continuum are due to reflection by dust, then we might expect the reflected light to be polarized.  \cite{LR13} find NIR polarization up to $\sim5$\% on scales of $<3$\arcsec, with inferred intrinsic polarization of $\sim12.5$\%.  But these measurements are dominated by the ENLR bicone and their observations do not cover the spatial scales of the rays observed by {\it HST}.

With the crepuscular rays seen on Earth at sunrise or sunset, the rays appear perpendicular to a nearly plane-parallel obscuring medium, but the clumpiness of the layer permits partial transmission of light.  This suggests that the situation in IC 5063 could be similar since the large-scale dust lanes should exist primarily in a plane.  The AGN might therefore shine through a clumpy large-scale disk in the galactic plane which forms large-scale dust lanes.  Although this explanation is simple, it introduces geometric issues: since the bright rays require scattering on projected $\sim11$-kpc scales, an extremely large plane-parallel extent of the scattering medium should be required.  We therefore expect a more isotropic distribution of the scattering dust.

\subsection{A Shadow Cast by the Torus?}

The orientation of the bicone relative to the galactic plane suggests a striking possibility: if the excess light at bright angles results from reflection of AGN emission off of diffuse galactic dust, then the dark rays could be the projected shadow of nuclear dust (possibly the torus), cast by AGN light.  This shadow prevents the subsequent reflection of AGN emission towards the observer at larger radii.  The dark and light pattern is therefore analogous to ``crepuscular rays" seen in the form of sunbeam patterns cast by clouds when the sun is obscured by the horizon.  Possible origins of the dust in the bright rays include expulsion by an AGN wind and loss during a merger that formed IC 5063. 

The bright rays in the middle of the dark sectors may pose a problem for an explanation based on a uniformly obscuring torus.  Extended X-ray emission is observed at large angles from the bicone axis, and provides evidence for clumpy torii full of gaps or holes \citep{Fabbiano18}.  But the bright rays in IC 5063 are narrow and well-defined, particularly in the SW.  A warped, clumpy torus could explain this configuration, since the column density could be much higher at the inner and outer rings of the torus than in the transition zone.  This also explains the lower intensity of the bright central ray, relative to the bounding edges.  The specific case of shadowing by a warped, clumpy torus is depicted in Fig. \ref{fig:cartoons}, upper-right.  Such sharp, well-defined edges in a scattering cone requires a nearly edge-on viewing angle, which helps explain why such a phenomenon has not previously been observed in other nearby AGN.  

The opening angle of unobscured optical emission in IC 5063 would be wide, $\sim137\degr$, implying that any shadowing torus must be thin.  Note that this angle is much larger than the angle subtended by the Seyfert-like emission in the bicone ($\sim25\degr$) that requires a high ionization parameter from incident AGN radiation, or possibly shock excitation by the narrow, bifurcated outflow.

\cite{Mingozzi19} use MUSE observations of the [\ion{S}{2}] doublet ratio to infer electron density $n_e\sim10^{3}\,\rm{cm}^{-3}$ in a $r\sim350$\,pc strip perpendicular to the bicone axis.  If the multiphase gas includes neutral gas at similar density, this implies a Compton-thick line-of-sight column density $n_{\rm H}\sim10^{24}\,\rm{cm}^{-2}$ at the midline of the dark rays. 

\subsection{Alternate Scenarios?}

The bicone orientation towards the galactic plane also suggests that the dark rays are unlikely to be due to the destruction of dust by radiation from the ionization cone.  In principle outflowing gas escaping from the galaxy is sufficiently hot \citep[$\simgreat10^8$\,K][]{Mukherjee18}, but it is not immediately clear whether the outflow dynamics are capable of maintaining the conical pattern.  Fig. \ref{fig:cartoons}, lower-left, depicts an extreme case where dust within the dark sector has been destroyed or vacated by hot outflows, and hence cannot reflect incident radiation from the AGN.

Excess emission relative to the elliptical models is concentrated in the disk, and in ``bright rays" immediately parallel to the dark rays.  The disk excess could be due to dusty reflection or excess starlight, but a stellar explanation for the bright rays would require an extreme X-shaped configuration in IC 5063.   

\begin{figure} 
\centering
\noindent
\begin{overpic}[width=0.9\linewidth]{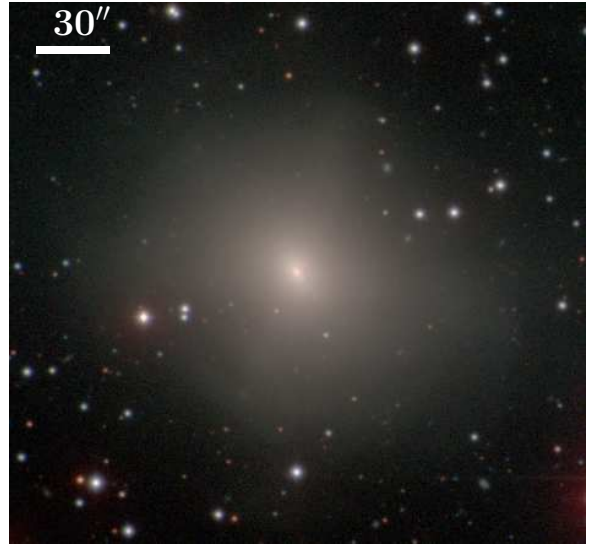}
	\put(4.7,85){\color{white}\rule{0.973cm}{0.1cm}}
	\put(7.5,89.5){{\parbox{1cm}{%
			\color{white}\bf\large 30\arcsec
			}}}
\end{overpic}\par
\caption{Carnegie-Irvine Galaxy Survey composite color image of IC 3370 \citep{Ho11CGS1}, which displays large-scale X-shaped features associated with post-merger stellar shell structure.}
\label{fig:ic3370}
\end{figure}

An `extreme stellar' scenario is depicted in Fig. \ref{fig:cartoons}, lower-right.   A merger with a smaller galaxy can produce extended concentric stellar shells which have an X-like appearance when viewed perpendicularly to the smaller galaxy's line of motion \citep{Pop18}.  IC 3370 displays such features, for example (Fig. \ref{fig:ic3370}; see also Figs. 7.82 and 12.6 respectively in the online editions of \citealt{Ho11CGS1} and \citealt{Tal09}).  IC 5063 is a well known post-merger system \citep{Colina91} but displays no evidence for stellar shells in our model-subtracted {\it HST} images.  This scenario might be preserved if a second minor merger disrupted the shells, or if phase mixing in an evolved merger has largely erased radial brightness stratification.

Peanut-shaped bulges can also produce extreme X-shaped stellar structures when viewed edge-on \citep{Parul20}.  In NGC 1175 ({\it HST} Program 15445, PI: Keel), these features reach comparable lengths ($\sim11$\,kpc).  Such structures are unlikely to subtend such large angles relative to the galactic plane as are seen in IC 5063, however (Parul, {\it private communication}), and IC 5063 is not a barred disk galaxy.  

\subsection{Constraints from Radial Profiles}

The radial profiles in Fig. \ref{fig:az} may provide clues to discriminating between the scenarios that involving dusty reflection or obscuration, although the interpretation is complicated by local structure such as dust and stellar lanes, particularly at small radii ($r\simless7\arcsec$).  In the bright outer rays ($\theta=[-1\degr,47\degr]$) the profiles eventually drop with increasing radius, and this drop might be dominated by geometric ($r^{-2}$) dilution of the AGN flux.  In the dark and bright central rays ($\theta=[11\degr,23\degr,35\degr]$, there is radial brightening until $r\sim15\arcsec$, beyond which the profiles flatten.

With the caveat that the ellipsoidal fits are better suited to emphasize azimuthal features and may be subject to spurious radial features, and that a proper treatment of dust scattering requires simulations beyond the scope of this letter \citep[e.g.][]{YMW84},  we note several apparent properties relevant to the models under consideration.  Firstly, if we ignore the structural complexity at small radii, then the radial brightening and flatness at large radii (i.e. if the profiles deviate from geometric dilution) could indicate recent AGN variability on light-crossing timescales, analogous to ENLR light echoes or {\it voorwerpjes} \citep{Keel15,Keel17}.  Subsequent simulations should address the impact of the dust density distribution, which could affect the absorbing column density at small projected radii, and the scattering efficiency at large projected radii.

Assuming the dust is optically thin to scattering, however, such an effect could be mimicked via limb brightening, which would depend upon the geometry of the dust distribution.  Another consideration introduced by the radial profiles is that if the central bright ray ($23\degr$ CCW from south) does not exceed the brightness of the dark rays ($11\degr, 35\degr$) at all radii, then the bright ray may not be easily explained by the ``warped, shadowing torus" model without additional considerations (e.g. radial dust distribution).  A proper treatment of the dust scattering is likely to require simulations beyond the scope of this paper, as has long been evident from bipolar reflection nebulae from other phenomena \citep[e.g.][]{YMW84}.

Variable obscuration caused by the transverse motion of clouds at small radii could produce the brightness inversion seen between the central ($23\degr$) and dark ($11\degr, 35\degr$) rays at around $r\sim15\arcsec$.  The angular size of the shadow and the light crossing time of the inversion zone imply that if the inversion is caused by the shadow of Keplerian material moving through the cone, it is likely cast by a large structural feature in the torus ($\sim500\,\kms$, $r_{\rm cloud}\sim3.5\times10^5\,R_{\rm Schwarzshild}$, $\sim0.2\,r_{\rm cloud}$ diameter) rather than a discrete cloud.

If the dark rays result from shadows cast by the torus or other nuclear dust lanes against a dust screen with effectively uniform angular symmetry, then a broad opening angle ($\simgreat137\degr$) is required, relative to the ionizing EUV radiation that typically penetrates through smaller angles and produces extended line emission in the bicone (e.g. NGC 3393, \citealt{Maksym16,Maksym17}).  Such differing illumination patterns should arise naturally if a modest column density can efficiently absorb the EUV at such large off-axis angles, but attenuates the optical and infrared light only modestly.  In Maksym et al ({\it in prep}), we conduct a detailed investigation of the geometry of the narrow line emission in IC 5063.

\section{Conclusion}

{\it HST} imaging shows that the extended near-infrared emission from IC 5063 displays large ray-like features relative to ellipsoidal models of the continuum, and these features extend to $\simgreat11$\,kpc.  The features are visible in filters covering 6868-9626\,\AA\ and 7165-8091\,\AA, and are oriented perpendicularly to the AGN bicone.

This configuration permits an intriguing explanation: that the dark rays originate in shadowing by the torus or a nuclear dust lane. Such shadowing is possible if a significant fraction of the near-infrared light in IC 5063 arises from AGN emission that is scattered by extended dust (which is roughly compatible with the the observed excess brightness).  If so, then the AGN torus may have a wide ($\simgreat137\degr$) opening angles at these wavelengths, as may AGN tori more generally.  The bright rays in the middle of the dark sectors may result from transmission by gaps in the obscuring material, such as may be possible with a warped, clumpy torus, or if the dust is concentrated in entrained filaments lifted by hot lateral outflows.

We have not entirely excluded alternate explanations like dust destruction or an encounter that produces an unusual configuration of X-shaped stellar orbits.  Except for the stellar encounter, all scenarios require scattering by extended dust and suggest a possible role for hot plasma vented perpendicular to bicone axis (such as in \citealt{Mukherjee18}, which suggests that such plasma is produced when the jets in IC 5063 heat and ablate the ambient ISM of the galactic plane).  The major question is whether this venting process preferentially displaces dust along the galactic axis, or destroys it, or concentrates it, or is effectively uniform on these spatial scales.  Further modeling is needed to investigate the effects from the dust properties, and from variable nuclear emission or obscuration.

We expect that deep near-infrared and optical polarization observations should be able to test whether the extended light outside the dark rays does indeed arise from dust scattering of AGN light.  New {\it HST} continuum observations of background galaxies at $\sim4000-5000$\AA\ would constrain the presence of an extended `dark cone' caused by entrained dust.  Deeper, wider {\it HST} observations might reveal signatures of stellar shells.  {\it JWST} may also be able to directly detect the presence of dust signatures, such as polycyclic aromatic hydrocarbons (PAHs).  Deep {\it Chandra} data may also be able to test alternate AGN-related explanations.  Deep MUSE spectroscopy of the bright rays may also be able to distinguish between scattered AGN light and starlight.


\acknowledgments

WPM acknowledges support by Chandra grants GO8-19096X, GO5-16101X, GO7-18112X, GO8-19099X.  Support for this work was provided by NASA through grant number HST-GO-15350.001-A from the Space Telescope Science Institute, which is operated by AURA, Inc., under NASA contract NAS 5-26555. 

 LCH was supported by the National Science Foundation of China (11721303, 11991052) and the National Key R\&D Program of China (2016YFA0400702).  MK was supported by the National Research Foundation of Korea (NRF) grant funded by the Korea government (MSIT) (NRF-2020R1A2C4001753).  This work was facilitated by Twitter discussions. Subvert the dominant paradigm.

%

\facilities{HST(ACS, WFC3, WFPC2)}








\end{document}